\documentclass[conference]{IEEEtran}
\IEEEoverridecommandlockouts
\usepackage[numbers]{natbib}
\setcitestyle{open={[},close={]}}

\usepackage{amsmath,amssymb,amsfonts}

\usepackage{graphicx}
\usepackage{textcomp}
\usepackage{xcolor}
\usepackage{float}
\usepackage{array}
\usepackage[font=footnotesize,labelfont=bf]{caption}
\usepackage{optidef}
\usepackage{subcaption}
\usepackage[utf8]{inputenc}
\def\BibTeX{{\rm B\kern-.05em{\sc i\kern-.025em b}\kern-.08em
    T\kern-.1667em\lower.7ex\hbox{E}\kern-.125emX}}

\usepackage{algorithm}
\usepackage{algpseudocode}
\usepackage{color, xcolor, colortbl}

\begin{document}

\title{Towards a Scalable and Trustworthy Blockchain: IoT Use Case\\}
\author{\IEEEauthorblockN{
Hajar Moudoud \IEEEauthorrefmark{1}\IEEEauthorrefmark{3},
Soumaya Cherkaoui  \IEEEauthorrefmark{1},
Lyes Khoukhi \IEEEauthorrefmark{2}
}

\IEEEauthorblockA{
\IEEEauthorrefmark{1} Department of Electrical and Computer Engineering, Université de Sherbrooke, Canada\\ 
\IEEEauthorrefmark{2} GREYC CNRS, ENSICAEN, Normandie University, France\\
\IEEEauthorrefmark{3} University of Technology of Troyes, France \\ 
\{hajar.moudoud, soumaya.cherkaoui\}@usherbrooke.ca, lyes.khoukhi@ensicaen.fr}
}

\maketitle
\begin{abstract}
Recently, blockchain has gained momentum as a novel technology that gives rise to a plethora of new decentralized applications ($e.g.$, Internet of Things (IoT)). However, its integration with the IoT is still facing several problems ($e.g.$, scalability, flexibility). Provisioning resources to enable a large number of connected IoT devices implies having a scalable and flexible blockchain. To address these issues, we propose a scalable and trustworthy blockchain (STB) architecture that is suitable for the IoT; which uses blockchain sharding and oracles to establish trust among unreliable IoT devices in a fully distributed and trustworthy manner. In particular, we design a Peer-To-Peer oracle network that ensures data reliability, scalability, flexibility, and trustworthiness. Furthermore, we introduce a new lightweight consensus algorithm that scales the blockchain dramatically while ensuring the interoperability among participants of the blockchain. The results show that our proposed STB architecture achieves flexibility, efficiency, and scalability making it a promising solution that is suitable for the IoT context.
\end{abstract}

\IEEEpeerreviewmaketitle

\section{Introduction}
\label{sec: Introduction}
With the widespread adoption of blockchain technology and global endorsement of cryptocurrencies, it is expected that transaction traffic will skyrock \cite{icc5}. According to \cite{r1}, the worldwide adoption of blockchain technology is expected to grow from 1.5 billion in 2018 to an estimated 15.9 billion by 2023; this will lead to an exponential growth in the number of transactions. Giving the inherent characteristics of nowadays blockchain ($i.e.$, extremely complex computing, low latency, and lacking dynamic scalability), handling the constantly growing number of transactions is becoming complex and very laborious. Blockchain sharding is a concept that was proposed to improve the blockchain scalability issue, which artificially partitions the workload of one single transaction procession to several members working in parallel. This separation brings about numerous advantages, including high flexibility, high throughput, and high scalability. The first proposed blockchain sharding solutions ($e.g.$, \cite{r2}, \cite{r3}, \cite{icc8}) deploy and on-chain sharding using Byzantine consensus protocol to improve the shards’ throughput and off-chain sharding moving transactions off the blockchain. Despite their efficiency and simplicity, these solutions have failed to meet the data validity required by nowadays applications \cite{icc10}. The reason is twofold: (1) with the increasing number of members working on a single transaction, a malicious member can compromise the work process, which may lead to security issues; and (2) any minor mishandling of a transaction may lead to its loss.

Today blockchain applications enable various industry verticals, such as supply chain management, healthcare, finance, and the Internet of Things (IoT). In particular, the blockchain may play a role in how IoT connected devices communicate and exchange information with each other. In this case, a blockchain transaction is defined as the basic communication primitive to exchange information among IoT connected devices. Handling a large number of transactions coming from these devices, particularly when they have heterogeneous security parameters, put the transaction reliability at risk, and introduces new security challenges. For instance, a compromised IoT device may deliver falsified information that will be stored permanently on the blockchain. This may impact the reliability of the transactions, which in turn affects the security of the blockchain.  

There have been several researchers have tried to address the challenges of using blockchain in an IoT context \cite{r23,icc4}. For instance, \cite{r4} and \cite{r5} used a distributed time-based consensus algorithm to ensure end-to-end security and ensure trustworthy transactions. Meanwhile, researchers in \cite{r6} and \cite{r7} have proposed a trustworthy blockchain to establish reliable and transparent data. However, the transactions verification coming from the outside world is challenging. In this context, oracles were proposed to verify transactions coming from off-chain applications. However, using a third party to verify the data may lead to a single point of failure, especially if the oracles are jeopardized \cite{icc9}.

To address these challenges, in his paper we present a scalable and trustworthy blockchain (STB) architecture; which provides scalable, trustworthy, transparent, and secure data exchange between unreliable IoT devices.

The main contributions of our paper can be summarized as follows: 
\begin{itemize}
	\item We design a secure blockchain architecture, called STB, which is a distributed and hierarchical architecture that is suitable for IoT applications. STB implements sharding and oracles to ensure efficient, scalable, and trustworthy data exchange among unreliable IoT devices.
	\item We propose a lightweight blockchain consensus that uses a decentralized trust method; which ensures secure deployment of the shards and prevents data loss. 
	\item We evaluate the performance of our architecture in terms of efficiency, latency, computing power, and scalability. The experiment results show that STB architecture can effectively be used to ensure a trustworthy, scalable, and efficient data exchange among IoT devices.  
\end{itemize}

The rest of this paper is organized as follows. Section II surveys some related work. Section III describes the architecture of the scalable and trustworthy blockchain (STB) for IoT and explains the lightweight blockchain consensus. Section IV presents a simulation-based case study while Section V concludes the paper.

\section{Related work}
\label{sec: Related works}
Blockchain technology is considered as the new solution to establish trust among unreliable entities. Due to the limited block size and block intervals ($i.e.$, the time needed to forge a new block) of traditional blockchain; the scalability and efficiency of the blockchain need to be further investigated. To overcome these problems, several blockchain networks were introduced to optimize these parameters \cite{r10} \cite{r11}. In \cite{r12}, Kouzinopoulos et al. proposed a new method to improve the block creation rate, an alteration to the way miners construct and reorganize blocks. In \cite{r13}, Lewenberg et al. proposed increasing the block size limit to improve the performance. However, increasing the block size or decreasing the block interval can lead to a stale block rate or double-spending vulnerability \cite{r14}. In \cite{r15}, Yang et al. proposed to only delegate blockchain consensus to a limited number of miners in the network. Although this solution decreased block intervals, it is prone to centralize since only a limited number of miners are responsible for the mining. 

Another approach to scale blockchain is to spread the workload on multiple blockchains simultaneously, this is referred to as blockchain sharding. To increase the blockchain scalability, multiple shards work in parallel. Consequently, this approach may lead to several security issues ($e.g.$, data loss, single shard attack). Particularly, a shard may be compromised and lead to data leakage \cite{r16} \cite{r17}. Besides the security issues, blockchain sharding can suffer from a wide range of security challenges, such as data interoperability or data validity. For instance, when a transaction is sent to the blockchain shards, each shard verifies a partition of the transaction, when all the sub-shards are verified, the cross-interoperability ($i.e.$, communication) is difficult. 

A major limitation of blockchain technology is its inability to interact with data outside of the chain ($i.e.$, outside world data) \cite{r18}. To overcome this issue, oracles were proposed to attest to facts in an effort to bring outside world data into the chain. In \cite{r20}, Nelaturu et al. proposed an oracle that verifies and transfers online data to the blockchain. However, online data could be maliciously altered or may be disclosed. In \cite{r19}, Ritzdorf et al. proposed a zero-knowledge oracle that authenticates data without disclosing full data. This oracle eliminates the need for a trusted third party and ensures data confidentiality. Although useful, their proposal have high computational costs.

To address the weaknesses of these existing solutions, we propose STB, a scalable and trustworthy blockchain architecture that ensure trust among IoT device in a distributed manner \cite{icc6}. STB architecture employs multiple shard chains processing transactions and one main chain to coordinate between the shards and ensure the transaction interoperability. Furthermore, we proposed using a decentralized oracle network to overcome the problem of single point failure and ensure a distributed trust among the participants of the network. And using a  using lightweight blockchain consensus, our architecture is less computing intensive in comparison with \cite{r20}\cite{r19}, making it suitable for IoT constrained device. 

\section{SCALABLE AND TRUSTHWORTHY
BLOCKCHAIN}
\label{sec: system}
STB is a distributed blockchain architecture that enables devices with constrained resources ($e.g.$, IoT sensors) to exchange data in a reliable manner \cite{icc10}. This section first describes STB’s overall architecture and then explains how the shards and oracles services are carried out.	
\subsection{Architecture}
As illustrated in Fig.1, this paper considers data exchange in an IoT environment where a large number of unreliable IoT devices are connected through a wireless network ($i.e.$, LTE, 4G, 5G, etc.) and accessible to the outside world \cite{bb}. We propose a scalable and trustworthy blockchain architecture composed of three layers, as follows:(1) IoT devices responsible for sending data that need to be verified and stored in the blockchain; (2) decentralized oracle network responsible for verifying the external data out of the chain and ensuring its reliability; and (3) the shards responsible for scaling the blockchain while keeping security guarantees.
\begin{figure*}[t]
	\centering
	\includegraphics[scale=0.48]{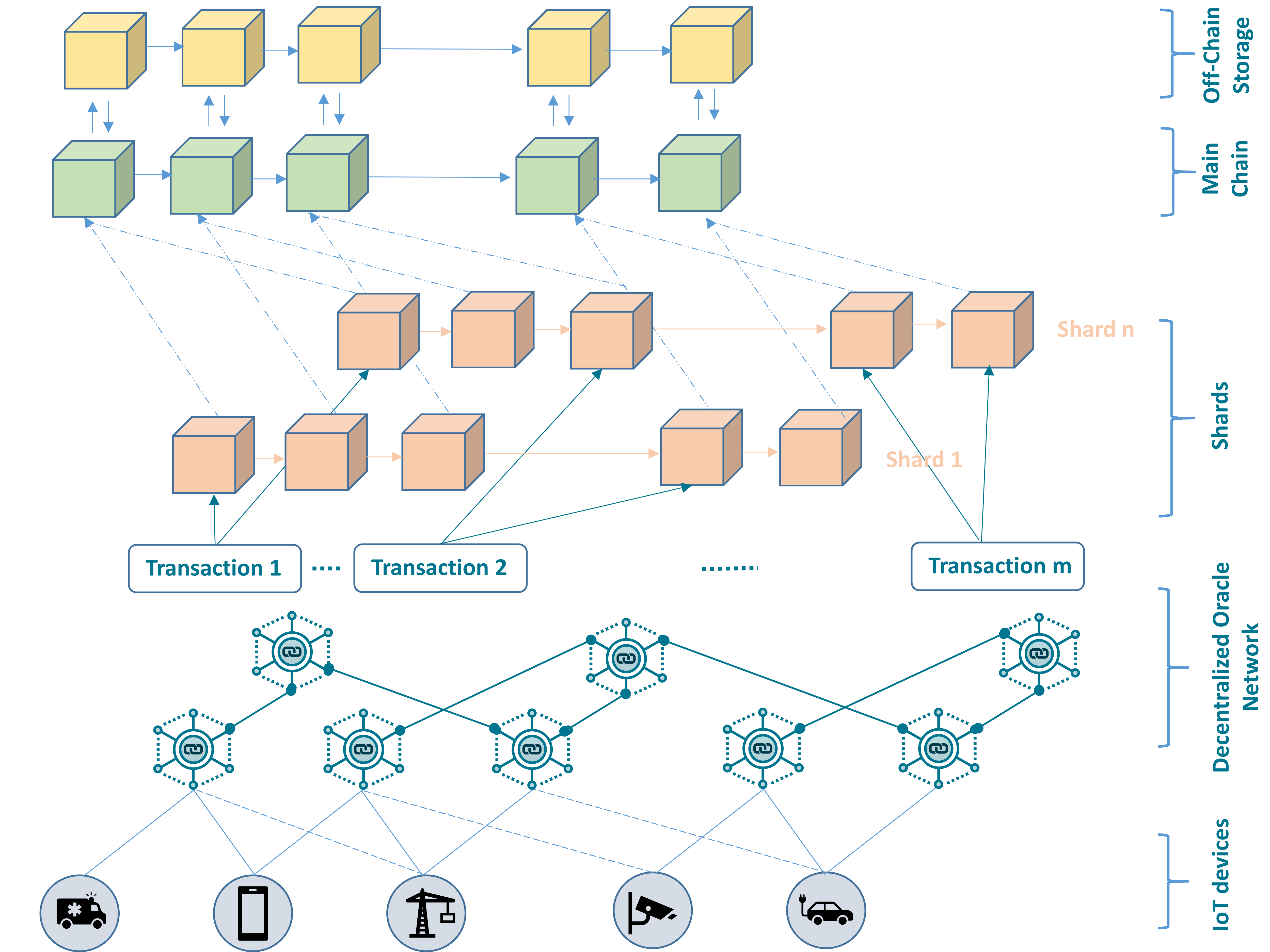}
	\caption{Scalable and Trustworthy Blockchain architecture}
	\label{fig:archi}
\end{figure*}
\subsubsection{Shared Blockchain} 
The scalability has been a core problem in the integration of blockchain within the IoT. This is because the IoT connects a large number of devices that generate big data. To enable horizontal scalability, blockchain sharding was proposed, it consists of partitioning each transaction into several shards and to process it independently. In this paper, we propose building clusters with multiple nodes ($i.e.$, miners) to process each shard in parallel.
\begin{figure}[t]
	\centering
	\includegraphics[scale=0.35]{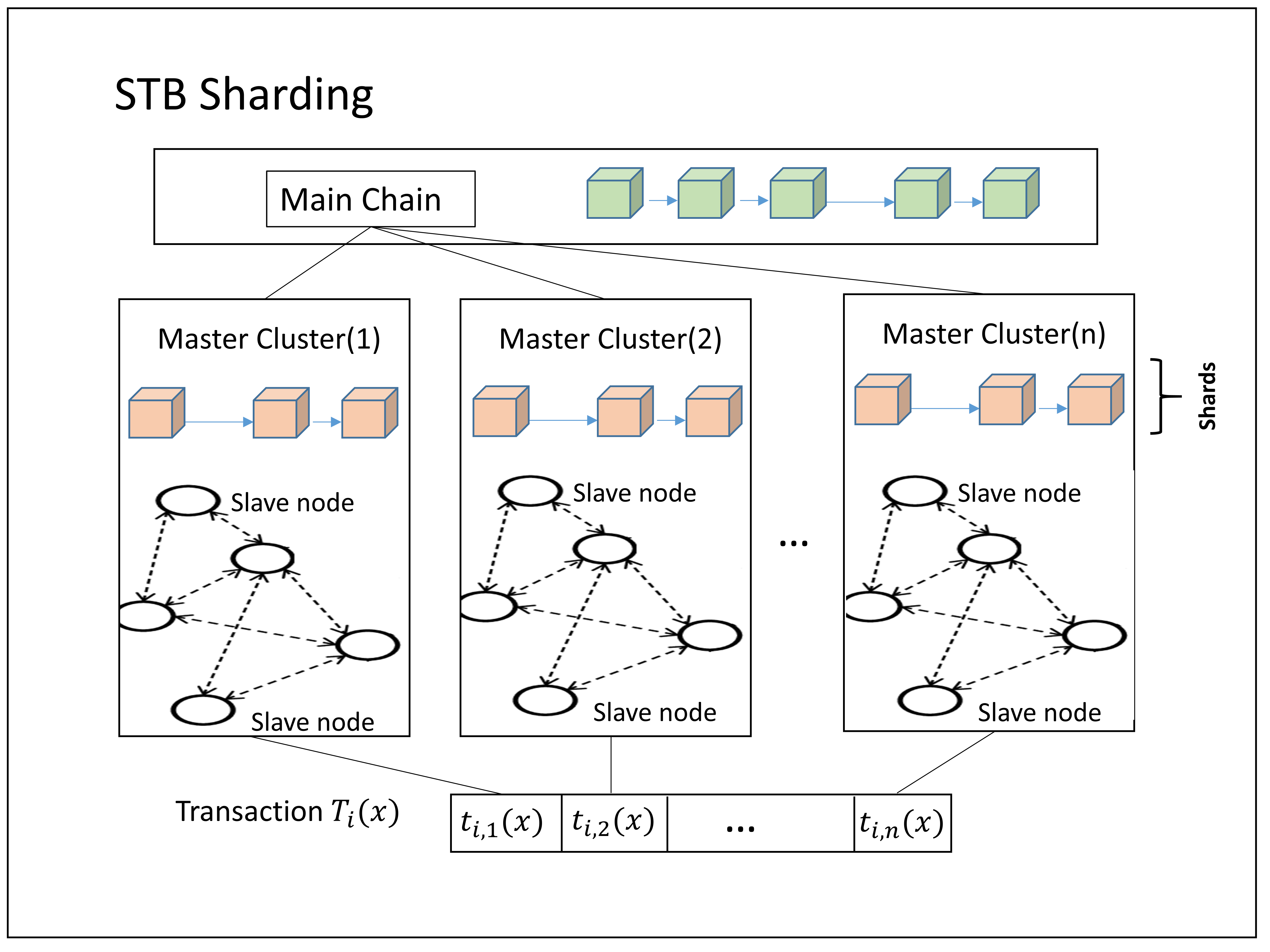}
	\caption{Scalable and Trustworthy Blockchain sharding}
	\label{fig:archi}
\end{figure}

Fig.2 shows the 3 main components of the STB sharding, as follows: (1) the main chain; (2) the master clusters'; and (3) slave nodes'. The master cluster contains $n$ slave nodes, that could be either honest or dishonest. Let $T_i(x)$ represents the $i_{th}$ transaction in a block. To enable the sharding, each transaction is divided into $n$ disjoint portion. All master clusters are responsible for verifying a portion of the transaction $t_{i,j}(x)$, where $i$ is the transaction number on a block and $j$ is the cluster number. All slave nodes could verify a transaction portion, outputs 0 or 1 indicates invalid and valid transaction, respectively. Let $n$ denote the total number of slave nodes in a master cluster $j$ that can tolerate up to $t \leqslant \frac{n}{3}$ dishonest node. We define a transaction verification process as follows:
\begin{equation}
T_{i}(x)= \sum_{j=0}^{n} \sum_{k=0}^{m} t_{i,j}(x) \times C_{j,k}(x)
\end{equation}
where, $C_{j,k}(x)$ is the trust value in a cluster$j$, proportional to a transaction portion $t_{i,j}$ and $\sum_{k=0}^{m} C_{j,k}=1$.

The transaction verification process has two possible outcomes: valid and invalid, each of which carries its own reward structure. As compensation for their efforts, the clusters are awarded ($R$) whenever they verify a transaction correctly. In the case of correct verification, a cluster reward is as follows.
\begin{equation}
R(T_{i,j}) =\frac{R(t_{i,j}(x))}{R(T_{i,Total}(x))} \times C_{j,k}
\end{equation}
where, $R(t_{i,j}(x))$ and $R(T_{i,Total}(x))$ represents, respectively, the reward for a valid transaction portion and total reward for a valid transaction. 
Note that the cluster trust value is only used to determine rewards and penalties, and does not necessarily correspond to the cluster output. Consequently, we enable secure sharding, prevent data loss, and incentive clusters to behave honestly. 
\subsubsection{Peer-To-Peer Oracle network}
We propose using a Peer-To-Peer (P2P) oracle network to verify the data queries and authenticate its source. Blockchain cannot access external data of the network. This is where the blockchain oracle interferes, it’s a service provider (trusted third party) that verifies the data authenticity. 	However, trusting a single third party may lead to providing corrupt or inaccurate data. To this end, we propose using a P2P oracle network that ensures the truth value of IoT retrieved data. We assume each received data sent from an IoT device could be either valid $V$ or false $F$. There is $m$ oracle in the P2P network, only $n$ oracle verify the data. For each oracle $o$  $\in$ [n,m] has a $q$ probability that data $d$ is correct about a given proposition.
\begin{equation}
O_i(d)=
\begin{cases}
q \quad  \text{when data is valid}\\
1-q \quad \text{when data is false} 
\end{cases}
\end{equation} 
The value of $o_i(d)$ is independent of $o_j(d)$ for all $i \neq j$. In other words, each oracle's trust values are independent of other oracles in the network trust values. Furthermore, oracles need to place a deposit to participate in a random-chosen verification process. First, the oracle submits its verification probability $q$ $\in$ [0,1] about a data $d$. Then, this probability is applied to a trust weight $W$ $\in$ [0,1] that is, informally, the parameter within the network that verifies the input data within the network's hidden layers. Formally:
\begin{equation}
W_{i}=\frac{\alpha_i}{\beta_i}
\end{equation}
where $\alpha_i$ is the sum of corrected verification performed and $\beta_i$ is the total number of verification performed. 
\begin{equation}
V(d)=\sum_{i=0}^{n}o_i(d)\times W_{i}
\end{equation}
Once a data verification process $V(d)$ has accumulated sufficient verifiability during a maximum of a period of time $\delta(t)$, it is decided. This period of time is a fixed value decided by network operators. The verification process has three possible outcomes:1) valid ($V$), if $V(d)$ value to be strictly positive, 2) false ($f$), if $V(d)$ value to be strictly negative, and 3) undefined ($U$) if $V(d)$ value is equal to zero. In this last case, the data verifiability process is only assigned to the oracle with the highest trust weight. 

Broadly speaking, oracles are rewarded when they participate in a verification process and their verification probability matches it\cite{icc2}. Conversely, those who gave incorrect verification are penalized. In the case of undetermined outcomes, oracles receive no rewards or penalties. As argued in the paper, the proposed data verification process incentives the oracles to behave honestly on the validity of data \cite{icc1, icc3}.
\subsection{Design Components}
The transaction verification procedure of our STB consists of the following (1) Initialization; (2) Consensus; and (3) Reward. This procedure starts with the initialization and then proceeds in periods, where each period consists of several iterations of consensus and finally the rewarding phase. We now explain each component in more detail. 
\subsubsection{Initialization} The initial set of participants ($i.e.$, nodes) are invited to provide a deposit ($stake$) to validate or endorse a transaction. That is, a node is given the chance to verify a transaction chosen uniformly at random from the unverified transaction pool. The deposit is placed before the verification process. Because the nodes are grouped in clusters, the outcome of the transaction verification reward is a function of the sum of the total transaction verification reward weighted by the cluster value in the cluster and the deposits. 
\subsubsection{Consensus}
Once nodes of each cluster are done with the initialization step, they wait for an IoT device to submit their transactions. The transaction is partitioned uniformly on $n$ partition. Each transaction partition is sent to a cluster that forwards the transaction to corresponding nodes. These clusters run an intra-cluster consensus protocol to approve the transaction partition. The inter-cluster consensus is a protocol where each node places a portion of its deposit in their confidence of their verification being valid. 

Meanwhile, it is challenging to build a consensus that processes transactions and ensures interoperability among the shards. Sharding consensus could lead to possible attacks, and thus without an efficient and secure cross-shard consensus data may be compromised and the blockchain interoperability may be lost. To solve this problem, we propose in this article an intra-cluster lightweight consensus to approve a transaction and add it to its ledger. Algorithm 1 illustrates the logic of this consensus. In STB we have the main chain and shard chains that accommodate $n$ sub-chains(shard). This consensus takes as input all partitions of a transaction $T_{i,j}$ and returns a new block if the transaction verification is valid; otherwise, it returns false. This intra-shard consensus will be invoked at the moment when all the collaborators (master clusters) try to forge a new block into the ledger. A transaction $T$ is partitioned equally among $n$ collaborators. First, we verify that the transaction has been verified by the P2P oracle network. We then calculate the transaction verification process according to Eq.1. If the output is superior to $\frac{n}{3}$, we then accept the transaction; otherwise, it is considered false. Finally, if the transaction is verified, we then forge a new block and update the ledger. 
\begin{algorithm}
	\hspace*{\algorithmicindent} \textbf{Input}  {$\forall t_{i,j}(x) \in T$, $O_i \in O$, $C_{i,j} \in C$;}\\
	\hspace*{\algorithmicindent} \textbf{Output} {$T$, $B_{i+1}$;}
	\textbf{Initialize:} \\
		\begin{algorithmic}[1]
\State $T$: A transaction from the unverified transaction pool;
 \State $n$: The number of master clusters;
\State $t_{i,j}$: A transaction partition;
\State 	$O_i$: An oracle form the P2P network;
\State 	$C_{i,j}$: A master cluster trust value;
	\If{$T.signature==O_i.sigature$}
	
	\While{$i \leqslant n$}
	
		\While{$C_{i,j} \in C$} 
		\State $T(x)= t_{i,j} \times C_{i,j} + T(x)$
		\State $W= W+C_{i,j} $	
	\EndWhile
	\EndWhile
\If{$W(x) \neq 1$}
\State Break;
\EndIf
\Else
	\If{$T(x) \leqslant \frac{n}{3}$}
		\State AddBlock($B_{i+1}$)
		\State $Return$ $Valid$ $Transaction$
	\EndIf
	
	\State $Return$ $False$ $Transaction$
\EndIf
\State $Break$

\end{algorithmic}
	\caption{Inter-Shard Consensus Algorithm.}	
\end{algorithm}

\subsubsection{Reward}
\begin{figure}[t]
	\centering
	\includegraphics[scale=0.32]{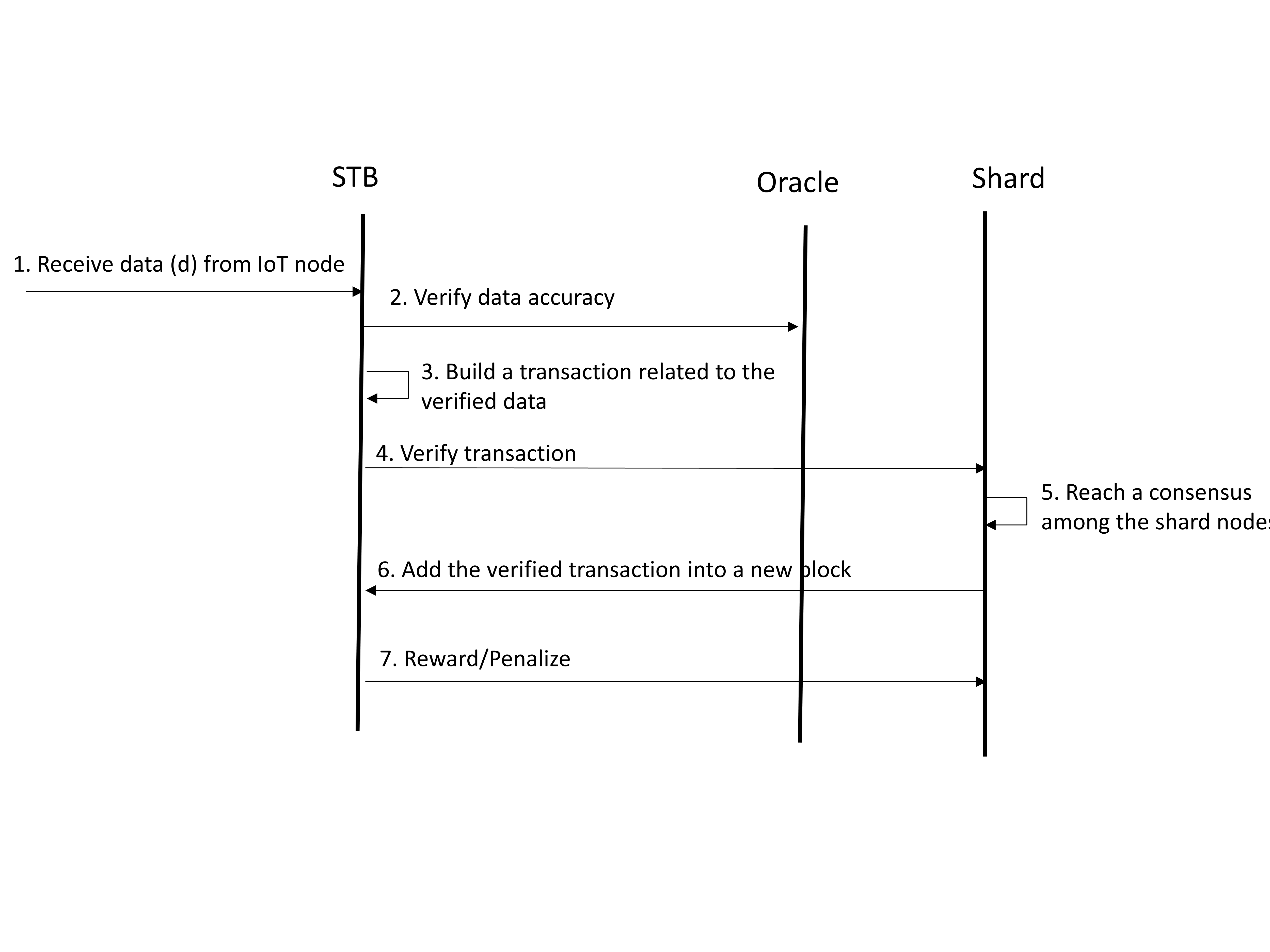}
	\caption{Adding data to the STB}
	\label{fig:archi}
\end{figure}
Broadly speaking, nodes ($i.e.$, miners) are rewarded for transaction verification in which they took part. Conversely, those who provided false verification are penalized. The steps of reward/penalize are depicted in Fig.3. Let $r_i$ denote the reward amount that a node $i$ used to verify that a transaction partition is equal to $0$ or $1$. In the case of a valid verification, the reward is as follows: 
\begin{equation}
r_i=\frac{R(T_{i,j})}{Total} + \alpha_i
\end{equation}
where, $R(T_{i,j})$ is the master cluster reward, $Total$ refers to the total number of the cluster node participating in the verification process, and $\alpha_i$ node deposit. A node's reward is equal to his participating share in the transaction verification process. For instance, if a cluster has a reward equal to 1000, and the number of participant nodes is equal to 100, the reward is distributed equally and each node will receive 100 as a stake. In the case of false verification, the node is penalized and the reward is deducted from its deposit.

\section{Evaluation}
\label{sec: evaluation}
This section presents the evaluation STB architecture. We compare our STB to other prominent works in terms of scalability, computing power, and latency \cite{icc4}. 
We consider several variant of the proposed architecture:(1) STB, which is a blockchain network that implements all of our proposed components and the proposed lightweight consensus; (2) Bitcoin \cite{r21}, which is a basic blockchain network that uses Proof-of-work(PoW)consensus; and (3) Quarkchain \cite{r22}, which is a blockchain network that implements the clustering feature and employs boson consensus- a blockchain sharding consensus to scale the network. The simulations were conducted on a laptop with 2.2 GHz Intel $i7$ processors and 16GB of memory.
\begin{table}[t]
	
	\begin{center}
		
		\caption{SIMULATION SETUP PARAMETER}\label{tab:2}
		
		\begin{tabular}{c | c}
			\hline
			\hline
			\textbf{Parameters} & \textbf{Value}\\
			\hline
			Oracles & 3\\
			IP & localhost\\
			Master clusters &  8\\
			Slave nodes & 10 \\        
			Maximum period of time $\delta(t)$ & 10$s$\\
			PoW difficulty & 4\\
			\hline
			\hline
		\end{tabular}
		
	\end{center}
\end{table} 
\begin{figure}[t]
	\centering
	\includegraphics[scale=0.48]{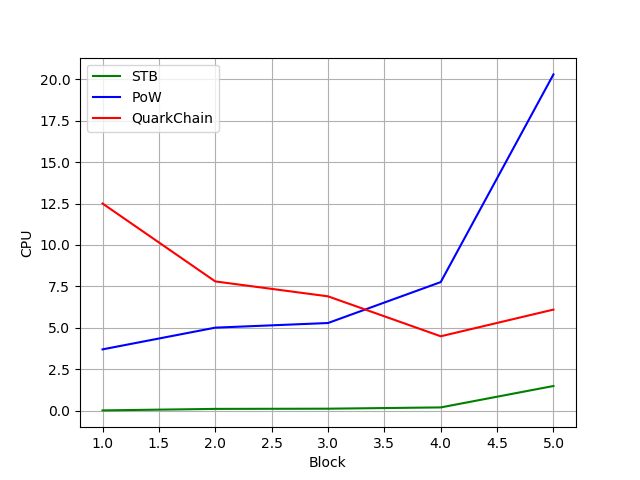}
	\caption{Scalability comparison}
	\label{fig:archi}
\end{figure}

To examine the scalability of different architecture, we evaluate the computing power needed to add one block to the chain. Fig. 4 depicts the results. Because Quarkchain uses the sharding concept, we can observe that the computing power needed to add blocks decreases with time. This happens because the block workload is separated among 8 shards. We clearly observe that the proposed STB architecture is less computing intensive than others. This happens because the proposed consensus algorithm does not have a difficulty level that limits block generations.

In Fig. 5, we evaluate the CPU in a function with time needed to add 10 blocks to the chain. In this experiment, we only compare STB with Quarkchain. This because, the time needed to add one block using Pow is much higher than other architecture. We observe that the local max, which point on the maximum CPU needed to forge blocks, of STB is less higher than Quarkchain. This is because, STB implements a distributed trust model that partition the work load needed to add blocks on several master clusters. These clusters in their turn partition the work load on the slave nodes that work in parallel.

In Fig. 6, we compare the latency of variant blockchains. As evidenced by the figure, the PoW latency increases with the number of blocks to add in the BC. Meanwhile, the time needed to add block ins Quarkchain is a fixed value $10s$ and $0.1s$ for the STB. The results demonstrate the efficiency
of STB, in terms of scalability, computing power, and latency in comparison with other architectures.
\begin{figure}[t]
	\centering
	\includegraphics[scale=0.48]{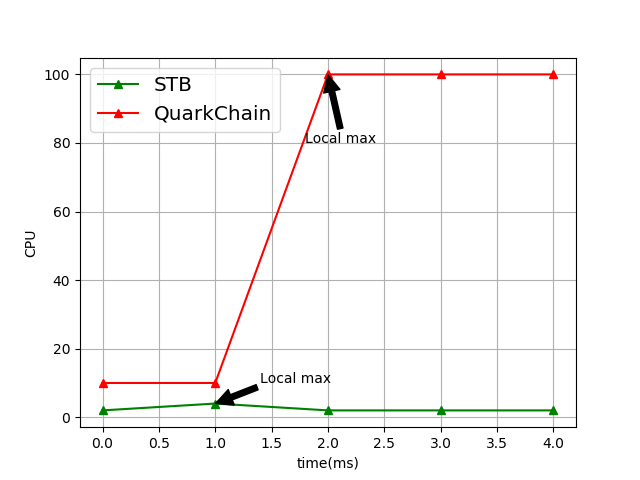}
	\caption{Computational power comparison}
	\label{fig:archi}
\end{figure}
\begin{figure}[t]
	\centering
	\includegraphics[scale=0.48]{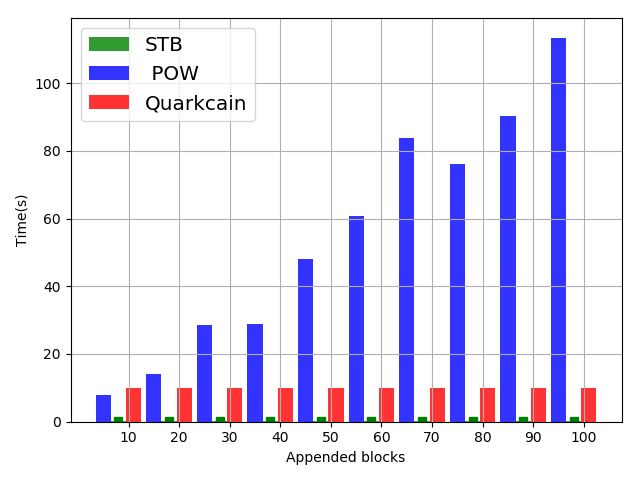}
	\caption{Latency comparison}
	\label{fig:archi}
\end{figure}

\section{Conclusion}
\label{sec: Conclusion}
This paper proposed, STB, which is a scalable and trustworthy blockchain architecture that is suitable for the IoT. Because of the large number of IoT  unreliable and constrained device, the integration of the blockchain with the IoT is not straightforward. To this aim, we first design the distributed and hierarchical blockchain architecture. We then describe the 2 main components of the chain, each of which is responsible for scaling the blockchain and verifying the reliability of information before storing them into the chain. Finally, we propose lightweight blockchain consensus that is suitable with IoT constrained device. The simulation results demonstrate that out architecture outperforms existing blockchain in terms of flexibility, scalability, and latency. 
\section*{acknowledgement}
\label{sec:acknowledgement}
The authors would like to thank the Natural Sciences and Engineering Research Council of Canada, as well as FEDER and GrandEst Region in France, for the financial support of this research.

\label{Related work}
\bibliographystyle{IEEEtran}
\bibliography{./references}


\end{document}